%% file: main.tex
\documentclass[sigconf, nonacm]{acmart} 

\AtBeginDocument{%
  }

\setcopyright{acmlicensed}
\copyrightyear{2026}
\acmYear{2026}
\acmDOI{XXXXXXX.XXXXXXX}

\usepackage{amsmath}    
\usepackage{booktabs}   
\usepackage{multirow}   
\usepackage{graphicx}
\usepackage{array}
\usepackage{float}

\begin{document}

\title{When LLM-Based User Profiling Adds Value in Production Streaming Recommendation}

\author{Milad Sabouri}
\orcid{0009-0009-8219-5596}
\affiliation{%
   \institution{DePaul University / Comcast Technology AI}
   \city{Chicago}
   \state{IL}
   \country{USA}}
\email{msabouri@depaul.edu}
\email{milad_sabouri@comcast.com}

\author{Neeraj Sharma}
\orcid{0009-0008-7940-1523}
\affiliation{%
   \institution{Comcast Technology AI}
   \city{Sunnyvale}
   \state{CA}
   \country{USA}}
\email{neeraj_sharma@comcast.com}

\author{Sardar Hamidian}
\orcid{0009-0007-5377-0374}
\affiliation{%
   \institution{Comcast Technology AI}
   \city{Washington}
   \state{DC}
   \country{USA}}
\email{sardar_hamidian@comcast.com}

\author{Shaghayegh Agah}
\authornote{Corresponding author.}
\orcid{0000-0002-7063-9251}
\affiliation{%
   \institution{Comcast Technology AI}
   \city{Sunnyvale}
   \state{CA}
   \country{USA}}
\email{shaghayegh_agah@comcast.com}

\begin{abstract}
Personalized recommendation depends critically on how user representations are constructed from historical behavior. Two paradigms have emerged for constructing semantic user profiles in content-based recommendation. First, aggregate methods derive user representations as numerical aggregates of semantic item embeddings. Second, LLM-based methods generate natural-language summaries of user preferences and encode them through a text encoder. Each paradigm can be combined with temporal disentanglement of recent versus historical behavior. LLM-based profile generation is significantly more expensive than aggregate approaches, raising the question of when this additional cost is justified. We present a systematic comparison of four semantic user-profiling strategies, factorially crossed across representation type and temporal handling, evaluated on a real-world production dataset. The comparison reveals how these strategies differ across user behavior types, across both accuracy and beyond-accuracy dimensions of recommendation quality, and across the temporal-window setting that governs the disentanglement. 

\end{abstract}

\ccsdesc[500]{Information systems~Recommender systems}

\keywords{Semantic User Profiling, LLMs, Recommender Systems}

\maketitle

\input{body}

\bibliographystyle{ACM-Reference-Format}
\bibliography{ref}

\end{document}

%% file: body.tex
\section{Introduction}
\label{sec:introduction}

 A central design choice in content-based recommendation is how to construct the user profile, which is the compressed representation of a user's historical preferences against which candidate items are ranked at retrieval and scoring time. Profile design strategy shapes both the accuracy of the resulting recommendations and the operational properties of the recommender pipeline, including the cost of profile generation, the cadence at which profiles are refreshed, and the engineering complexity of maintaining the system in production.

Two paradigms have emerged in the literature for constructing semantic user profiles in content-based recommendation. Aggregate methods derive user representations as numerical aggregates of semantic embeddings of items the user has interacted with, with simple centroid-based pooling as the canonical example~\cite{lops2011content}. More recent work has proposed LLM-based methods that generate user profiles as natural-language summaries of preferences, which are subsequently encoded into vector representations through a text encoder~\cite{sabouri2025towards, lubos2024llm, ma2024xrec}. Each paradigm can be combined with a temporal component that disentangles short-term and long-term user behavior through attention-based fusion of segment-level representations~\cite{bahdanau2014neural,waswani2017attention}. Prior work has demonstrated cases where LLM-based profiling outperforms aggregate methods on public e-commerce benchmarks~\cite{sabouri2025towards}; whether this advantage extends to production streaming data, with its different scale, user behavior heterogeneity, and ranking conditions, has received less characterization.

Production streaming environments differ substantially from typical academic evaluation settings, where candidate catalogs are much larger, user histories are deeper, and consumption behavior is more heterogeneous, ranging from stable preference reinforcement to exploratory engagement. At the same time, LLM-based profiling introduces nontrivial computational, latency, and operational considerations compared with aggregate profiling. These realities raise a key question: when do the benefits of LLM-based profiling justify this increased complexity, and when are simpler aggregate methods more effective?
 
We address this question through a systematic offline evaluation on a sample from a production streaming platform. We compare four semantic user-profiling strategies in a 2$\times$2 design across two axes: representation type (aggregate item-embedding vs. LLM-generated summaries) and temporal handling (full-history profiles vs. temporally disentangled short- and long-term profiles fused via attention). This design allows us to examine whether LLM-based representations outperform aggregate methods, whether temporal disentanglement improves performance in either setting, and whether these effects vary by user behavior type.

Two findings emerge from our analysis. First, the relative effectiveness of LLM-based and aggregate semantic user profiling is segment-conditional. Aggregate methods dominate among habitual users, while LLM-based methods produce accuracy gains among exploratory users whose future consumption diverges from their past behavior. Second, LLM-based profiling exhibits counterintuitive recommendation behavior. Although individual recommendation lists become modestly more diverse, the system as a whole concentrates on a smaller and more popular subset of the catalog, with measurable reductions in catalog coverage and recommendation novelty. We additionally characterize the operating range of the temporal-window hyperparameter and discuss implications for deployment. Taken together, these findings suggest that simpler aggregate methods remain a robust default. The additional cost of LLM-based profiling is justified selectively, on user segments where it produces measurable accuracy gains.

\begin{table*}[t]
  \caption{\small The 2$\times$2 design space of semantic user-profiling strategies compared in this work. Rows correspond to representation type; columns correspond to temporal handling.}
  \label{tab:design_grid}
  \small
  \begin{tabular}{@{}p{0.18\linewidth} p{0.36\linewidth} p{0.36\linewidth}@{}}
    \toprule
    & \textbf{Holistic} \newline (single profile over full history) & \textbf{Temporal} \newline (short / long-term split, attention-fused) \\
    \midrule
    \textbf{Aggregate} \newline (SBERT centroid of items) & \textbf{Centroid (C)} \newline Single centroid of SBERT embeddings over the full interaction history. & \textbf{TemporalCentroid (TC)} \newline Short-term and long-term SBERT centroids combined via attention. \\
    \addlinespace[2pt]
    \textbf{Narrative} \newline (LLM-generated NL summary, SBERT-encoded) & \textbf{Narrative (N)} \newline Single LLM-generated natural-language summary of user preferences, SBERT-encoded. & \textbf{TemporalNarrative (TN)} \newline Two LLM-generated natural-language summaries (recent and full history), SBERT-encoded and fused via attention. \\
    \bottomrule
  \end{tabular}
\end{table*}


\section{Related Work}
\label{sec:related_work}

\paragraph{Aggregate user profiling}
\label{sec:related_aggregate}

Content-based recommendation has a long tradition of constructing user representations by aggregating item-level features~\cite{lops2011content}. With the rise of dense semantic embeddings, items became represented in continuous embedding spaces produced by language models, and user profiles were constructed as centroids or other pooling functions over these embeddings. Sentence-level embeddings, particularly those produced by Sentence-BERT~\cite{reimers-2019-sentence-bert}, became standard for content-based item representation. Variants such as weighted aggregation, recency-decayed pooling, and attention-based fusion have been explored as refinements to plain centroid aggregation.

\paragraph{LLM-based user profiling}
\label{sec:related_llm}

A more recent line of work proposes the use of large language models to generate user profiles as natural-language summaries of preferences, rather than as numerical aggregates of item embeddings~\cite{lubos2024llm, ma2024xrec}. The premise is that an LLM, given a structured representation of a user's interaction history, can produce a thematic summary that captures preference patterns at a higher level of abstraction than item-level pooling. This summary is typically encoded into a vector through a sentence encoder, allowing it to be consumed by the same downstream architecture used for aggregate profiles.

LLM-based profile generation has been explored in several contexts, including explainable recommendation~\cite{lubos2024llm, ma2024xrec}, sequential recommendation, and user understanding for personalization. The motivation in these works is that natural-language profiles offer interpretability and the potential for thematic generalization beyond observed items. Empirical evaluation has been conducted primarily on academic benchmarks, with the relative cost of LLM-based profile generation acknowledged but not deeply characterized in production settings. Closely related to our setting, recent work has proposed temporally disentangled LLM-based profiling that generates separate short-term and long-term natural-language summaries~\cite{sabouri2025towards}; we adopt the prompting, the idea of short and long term preference split, and attention-fusion patterns established in this line of work for the temporally disentangled LLM strategy in our comparison (Section~\ref{sec:temporal_narrative}).

\paragraph{Temporal user modeling}
\label{sec:related_temporal}

Modeling temporal aspects of user behavior is a long-standing thread in recommender systems research. Sequence-based models, including recurrent neural networks~\cite{tan2016improved}, self-attentive sequential recommenders~\cite{kang2018self}, and bidirectional transformers~\cite{sun2019bert4rec}, treat the user's interaction history as an ordered sequence and learn to predict the next item conditional on the sequence. These methods have demonstrated strong performance on next-item prediction tasks and have informed much subsequent work on temporal modeling in recommendation.

A complementary line of work decomposes user behavior into short-term and long-term components rather than modeling the sequence directly~\cite{zhu2017what, wang2019modeling}. Short-term components capture recent interests and behavioral shifts, while long-term components capture stable preferences across the full history. The two components are typically combined through attention-based fusion, with the relative weighting learned from data. The architectural patterns for attention-based fusion in this setting trace to broader work on attention mechanisms in deep learning~\cite{bahdanau2014neural, waswani2017attention}. Both temporally disentangled strategies in our comparison (TemporalCentroid and TemporalNarrative) build on this short-term/long-term framework, applied within the content-based ranking paradigm.

\paragraph{Production-scale evaluation}
\label{sec:related_production}

A small but growing literature addresses the gap between academic recommender evaluation and production deployment conditions. Industry-track and applied research have raised concerns about the candidate-pool sizes used in academic benchmarks, the absence of full-catalog ranking, the limited treatment of user heterogeneity, and the underrepresentation of beyond-accuracy metrics relevant to deployment such as catalog coverage and item novelty~\cite{vargas2011rank, castells2015novelty}. Segment-level analysis, in which evaluation metrics are computed separately for distinct user populations, has also been advocated as a tool for understanding where recommendation methods produce value and where they fail\cite{kim2025time}. The user-conditional analysis we present in Section~\ref{sec:results} draws on this body of work and applies it to characterize the deployment-relevant properties of LLM-based versus aggregate user profiling.


\section{Methodology}
\label{sec:methodology}

We compare four semantic user-profiling strategies for content-based recommendation, factorially crossed across two design axes. The first axis, \emph{representation type}, distinguishes profiles derived as numerical aggregates of item embeddings from profiles generated as natural-language summaries by a large language model. The second axis, \emph{temporal handling}, distinguishes profiles constructed over the user's full chronological history from profiles temporally disentangled into short-term and long-term components fused via attention.

The four resulting strategies are named \emph{Centroid} (C), \emph{TemporalCentroid} (TC), \emph{Narrative} (N), and \emph{TemporalNarrative} (TN). Their organization across the two design axes is summarized in Table~\ref{tab:design_grid}. All four strategies share a common downstream architecture: a user profile vector $\mathbf{p}_u \in \mathbb{R}^d$ is concatenated with an item embedding $\mathbf{e}_i \in \mathbb{R}^d$ and passed to a multi-layer perceptron (MLP) that produces a relevance score. Item embeddings are computed identically across strategies by encoding textual item metadata with a pretrained SBERT model~\cite{reimers-2019-sentence-bert}. The strategies differ only in the construction of $\mathbf{p}_u$, isolating user-profile design as the sole experimental variable. 

The four strategies compared in this work draw on established methodological traditions. The aggregate strategies (Centroid and TemporalCentroid) build on conventional centroid-based content profiling~\cite{lops2011content}; TemporalCentroid additionally applies attention-based fusion of short- and long-term user
segments~\cite{wang2019modeling, bahdanau2014neural}. The holistic LLM-based profile (Narrative) follows an established pattern of generating natural-language user summaries and encoding them into a shared embedding space. The temporally disentangled LLM-based profile (TemporalNarrative) combines these elements in an architecture previously evaluated on public e-commerce benchmarks~\cite{sabouri2025towards}. The contribution of this section is to organize these strategies into a 2$\times$2 design space and to isolate the construction of $\mathbf{p}_u$ as the sole experimental variable across the four conditions.

\subsection{Semantic user-profiling strategies}
\label{sec:notation}

Let $\mathcal{U}$ denote the set of users and $\mathcal{I}$ the catalog of items. For a user $u \in \mathcal{U}$, let $\mathcal{H}_u = \{(i, t) : i \in \mathcal{I},\, t \in \mathbb{R}\}$ denote the user's chronologically ordered interaction history, where each entry is a pair of an item $i$ and a timestamp $t$. Each item $i$ is associated with textual metadata $m_i$ comprising its title and description. The recommendation task is to score every item in the catalog for each user and return the top-$K$ highest-scoring items.

Scoring is performed as
\begin{equation}
  s(u, i) = f_\theta\!\left(\mathbf{p}_u \,\Vert\, \mathbf{e}_i\right),
  \label{eq:scoring}
\end{equation}
where $\Vert$ denotes vector concatenation, $f_\theta$ is a learned MLP with parameters $\theta$, $\mathbf{p}_u \in \mathbb{R}^d$ is the user profile vector produced by one of the four strategies described below, and $\mathbf{e}_i = \text{SBERT}(m_i)$ is the item embedding. The MLP, the item encoder, and the training and evaluation protocols are held identical across the four strategies.

\subsubsection{Centroid (C)}
\label{sec:centroid}

The Centroid profile is the arithmetic mean of SBERT\cite{reimers-2019-sentence-bert} embeddings of all
items in the user's interaction history:
\begin{equation}
\mathbf{p}^{C}_u = \frac{1}{|\mathcal{H}_u|} \sum_{(i, t) \in \mathcal{H}_u} \text{SBERT}(m_i).
\end{equation}
This strategy uses no temporal structure and produces no natural-language
artifact. It serves as the reference against which all other strategies
are compared.

\subsubsection{TemporalCentroid (TC)}
\label{sec:temporal_centroid}

TemporalCentroid computes two centroids over different temporal segments of the user's history. Given a short-term window parameter $\rho \in (0, 1)$,
the short-term segment $\mathcal{H}^{\text{short}}_u$ consists of the most recent $\rho \cdot |\mathcal{H}_u|$ interactions sorted chronologically.
The long-term segment is the user's full chronological history $\mathcal{H}_u$, not the complement of the short-term segment. This choice preserves the recent interactions in both segments, ensuring the long-term profile captures persistent themes anchored by all of the user's behavior.

Two centroids, $\mathbf{c}^{\text{short}}_u$ and $\mathbf{c}^{\text{long}}_u$, are computed as the mean SBERT embeddings of items in their respective
segments and combined into a single profile vector via attention:
\begin{equation}
\label{eq:tc}
\mathbf{p}^{TC}_u = \alpha_u \cdot \mathbf{c}^{\text{short}}_u + (1 - \alpha_u) \cdot \mathbf{c}^{\text{long}}_u,
\end{equation}
where $\alpha_u \in (0, 1)$ is a softmax-normalized scalar produced from the two centroids by a learnable single-layer attention parameterized by $\mathbf{W}_a \in \mathbb{R}^{1 \times d}$.

\subsubsection{Narrative (N)}
\label{sec:narrative}

Given the user's full interaction history, an LLM is prompted to produce a single textual description $T^{N}_u$ of the user's preferences. A simplified version of the prompt template is: ``\textit{Given the user's complete viewing history below, produce a concise summary of the user's overall preferences, emphasizing enduring themes and recurring patterns across the full history.}'' The summary is encoded into a vector through the same SBERT model used for item embeddings:
\begin{equation}
\mathbf{p}^{N}_u = \text{SBERT}(T^{N}_u).
\end{equation}

\subsubsection{TemporalNarrative (TN)}
\label{sec:temporal_narrative}

TemporalNarrative combines LLM-based representation with temporal disentanglement. Using the same short-term window $\rho$ and segment definitions as TemporalCentroid, the LLM is prompted twice: a short-term prompt eliciting recent interests and behavioral shifts, and a long-term prompt eliciting persistent preferences. Simplified versions of the two prompts are: ``Summarize the user's most recent viewing patterns, focusing on shifts in genre, tone, or content type'' (short-term) and ``Summarize the user's persistent preferences across their full viewing history, emphasizing themes that have remained stable over time'' (long-term).

Let $T^{\text{short}}_u$ and $T^{\text{long}}_u$ denote the two resulting summaries. Each is encoded via SBERT and the two embeddings
are combined through the same attention-based fusion as in TemporalCentroid:
\begin{equation}
\mathbf{p}^{TN}_u = \alpha_u \cdot \text{SBERT}(T^{\text{short}}_u) + (1 - \alpha_u) \cdot \text{SBERT}(T^{\text{long}}_u),
\end{equation}
where $\alpha_u$ is computed from the SBERT embeddings of the two summaries by the same $\mathbf{W}_a$ parameterization described in
Section~\ref{sec:temporal_centroid}.

\subsection{Shared infrastructure}
\label{sec:shared}

The four profiling strategies differ only in the construction of $\mathbf{p}_u$. All other components are held constant.

\textbf{Item encoder.} All item embeddings are produced by a frozen pretrained SBERT model applied to the concatenation of each item's title and description. Embeddings are precomputed and cached.

\textbf{Scoring head.} A two-layer MLP with ReLU activations consumes $\mathbf{p}_u \,\Vert\, \mathbf{e}_i$ and produces a scalar relevance score, as in Equation~\eqref{eq:scoring}. The MLP architecture is identical across strategies.

\textbf{Training objective.} Each strategy is trained with binary cross-entropy loss using negative sampling: each positive interaction in the training set is paired with uniformly sampled negative items, and the model learns to score positives above negatives. Optimization uses the Adam optimizer~\cite{kingma2014adam}. 

The four strategies thus differ exclusively in the function $\mathbf{p}_u(\mathcal{H}_u)$, supporting clean attribution of observed differences in evaluation metrics to semantic user-profile design choices rather than to confounds in candidate scoring or item representation.

\begin{table*}[t]
  \caption{\small Accuracy results at $\text{PCT} = 25\%$. All values are percentage change relative to the Centroid, which serves as the reference and is reported as ``---''. The $^{\ast}$ indicates the difference from the centroid is statistically significant at $\alpha = 0.05$. Statistical significance varies with segment and $K$; the segment-conditional pattern is most robustly significant at $K = 100$, while $K = 10$ results on the Explorer segment are at the margin of significance owing to the smaller segment sample size, as discussed in Section~\ref{sec:results_accuracy}.}
  \label{tab:headline}
  \footnotesize
  \setlength{\tabcolsep}{4pt}
  \begin{tabular*}{\textwidth}{@{\extracolsep{\fill}}l l rrr rrr@{}}
    \toprule
    & & \multicolumn{3}{c}{$K = 10$} & \multicolumn{3}{c}{$K = 100$} \\
    \cmidrule(lr){3-5} \cmidrule(lr){6-8}
    Segment & Strategy & Recall & NDCG & HR & Recall & NDCG & HR \\
    \midrule
    \multirow{4}{*}{All}
    & Centroid          & ---     & ---     & ---     & ---    & ---     & ---    \\
    & TemporalCentroid  & $+0.5$  & $+4.7^{\ast}$  & $+1.4$  & $-0.3$ & $+2.9^{\ast}$  & $+0.3$ \\
    & Narrative         & $-30.7^{\ast}$ & $-26.9^{\ast}$ & $-20.9^{\ast}$ & $-9.0^{\ast}$ & $-11.4^{\ast}$ & $-6.1^{\ast}$ \\
    & TemporalNarrative & $-21.6^{\ast}$ & $-15.2^{\ast}$ & $-14.2^{\ast}$ & $-9.0^{\ast}$ & $-11.5^{\ast}$ & $-5.0^{\ast}$ \\
    \midrule
    \multirow{4}{*}{Non-Explorer}
    & Centroid          & ---     & ---     & ---     & ---     & ---     & ---    \\
    & TemporalCentroid  & $+0.7$  & $+4.8^{\ast}$  & $+1.7^{\ast}$  & $-0.3$  & $+3.0^{\ast}$  & $+0.3$ \\
    & Narrative         & $-31.7^{\ast}$ & $-27.8^{\ast}$ & $-21.8^{\ast}$ & $-10.2^{\ast}$ & $-11.9^{\ast}$ & $-7.2^{\ast}$ \\
    & TemporalNarrative & $-22.4^{\ast}$ & $-15.8^{\ast}$ & $-14.7^{\ast}$ & $-10.2^{\ast}$ & $-12.3^{\ast}$ & $-5.9^{\ast}$ \\
    \midrule
    \multirow{4}{*}{Explorer}
    & Centroid          & ---     & ---     & ---     & ---    & ---     & ---    \\
    & TemporalCentroid  & $-7.4$  & $-6.0$  & $-10.2$ & $-0.7$ & $-0.2$  & $+0.5$ \\
    & Narrative         & $+18.7$ & $+31.5^{\ast}$ & $+14.8$ & $+9.5^{\ast}$ & $+12.8^{\ast}$ & $+6.2$ \\
    & TemporalNarrative & $+14.3$ & $+21.8$ & $+9.0$  & $+11.4^{\ast}$ & $+10.2^{\ast}$ & $+7.1$ \\
    \bottomrule
  \end{tabular*}
\end{table*}

\section{Experiments}
\label{sec:experiments}

We evaluate the four strategies through an offline experiment on a sample drawn from a production streaming platform. 

\subsection{Experimental setup}
\label{sec:setup}

\subsubsection{Dataset and sample}
\label{sec:dataset}

The dataset is sampled from a production streaming platform serving Movies, TV Shows, and Sports content. We evaluate on a random sample of 10{,}000 users. The catalog is filtered to items whose title and description are in English and contain sufficient text for meaningful encoding, yielding approximately $2.4 \times 10^{5}$ items. For each user, the interaction history is split chronologically into 80\% training, 10\% validation, and 10\% test. The split is performed per user to prevent within-user temporal leakage.

\subsubsection{Scope: content-based comparison}
\label{sec:scope}

The four profiling strategies compared in this work are content-based, deriving user representations from the textual content of consumed items. We deliberately exclude sequence-based and collaborative methods such as SASRec~\cite{kang2018self} and BERT4Rec~\cite{sun2019bert4rec} from the comparison. These methods address a complementary problem, namely next-item prediction over interaction sequences using collaborative signal, and would introduce confounds in signal source were they to be compared directly to content-based profiling strategies. Holding the broader recommendation paradigm fixed at content-based ranking allows the 2$\times$2 contrast across the design space introduced in Section~\ref{sec:methodology} to isolate the effect of user-profile design within that paradigm.

\subsubsection{User segmentation}
\label{sec:segmentation}
Profiling-strategy effectiveness may depend on whether a user's future consumption resembles their past behavior or departs from it. To study this conditioning effect, we group users into two segments and report results separately for each. The segmentation is post-hoc and is used for analytical purposes only; how an inference-time proxy for exploratory behavior might be constructed for deployment is discussed in Section~\ref{sec:routing_implications}.

A user is labeled an \textit{Explorer} if no item in their test history appears in their training history, and a \textit{Non-Explorer} otherwise. Non-Explorers exhibit habitual consumption that reinforces established preferences, while Explorers represent a regime in which past consumption has limited predictive power for future behavior. In our sample, 1{,}761 users are classified as Explorers and the remaining as Non-Explorers. Results are reported for the full population (All) and for each segment separately.

\subsubsection{Evaluation metrics}
\label{sec:metrics}
We report six metrics at $K \in \{10, 100\}$, organized into two families.

\textbf{Accuracy.} Recall@$K$ is the fraction of test-set items that appear in the user's top-$K$ list. NDCG@$K$ is the standard rank-aware variant. HitRate@$K$ is the fraction of users with at least one test-set item in the top-$K$.

\textbf{Beyond-accuracy.} Diversity@$K$ is the average pairwise cosine dissimilarity within each user's top-$K$ list, capturing within-list semantic spread~\cite{castells2015novelty}. Coverage@$K$ is the fraction of the full catalog that appears in any user's top-$K$ list, capturing cross-user catalog footprint. Novelty@$K$ is the mean self-information of recommended items, $\text{Novelty@}K(u) = \frac{1}{K}\sum_{i \in R_u^K} -\log_2 p(i)$, where $R_u^K$ is the user's top-$K$ recommendation list and $p(i)$ is the popularity of item $i$, measured as the fraction of training-set users with at least one interaction with $i$~\cite{vargas2011rank}. Lower Novelty indicates that recommendations are concentrated on more popular items.

\subsubsection{Short-term window sweep}
\label{sec:pct_sweep} 

The short-term window parameter $\rho$ governs the temporal partition in TemporalCentroid and TemporalNarrative (Section~\ref{sec:temporal_centroid}). We refer to this parameter as PCT, the percentage of the user's chronological history assigned to the short-term segment, and sweep $\rho \in \{0.15, 0.20, 0.25, 0.30\}$, reported throughout as percentages. Centroid and Narrative are PCT-invariant by construction and produce identical results across all four settings; their values appear as a reference point in PCT-related comparisons.

\subsubsection{Reporting and statistical testing}
\label{sec:reporting}
All metric values are reported as relative percentage change with respect to a within-segment, within-$K$ reference, in accordance with platform confidentiality requirements. Two normalization conventions are used. For cross-model comparisons in tables and figures, the reference is the Centroid value at the same segment and same cutoff $K$. For the temporal-window sensitivity analysis in Section~\ref{sec:results_pct}, each model's trajectory is normalized to its own value at $\rho = 0.25$ within the same segment, isolating the shape of each model's response to PCT from absolute level differences across models.

To assess statistical reliability, we apply the Wilcoxon signed-rank test~\cite{wilcoxon_1945_individual, hollander_2014_nonparametric} to pairwise comparisons of per-user metric values, independently for each metric, cutoff $K$, and segment combination at PCT~=~25\%. The test is non-parametric and makes no distributional assumptions on per-user metric values. Statistically significant differences at $\alpha = 0.05$ are indicated with an asterisk in Tables~\ref{tab:headline} and~\ref{tab:beyond_accuracy}. Because Coverage is a system-level metric computed across all users rather than per user, it is not subject to per-user significance testing.

\begin{table*}[t]
  \caption{\small Beyond-accuracy results at $\text{PCT} = 25\%$. All values are percentage change relative to the Centroid, which serves as the reference and is reported as ``---''. The $^{\ast}$ indicates the difference from Centroid is statistically significant at $\alpha = 0.05$. $^{\dagger}$ Coverage is a system-level metric and is not subject to per-user statistical testing.}
  \label{tab:beyond_accuracy}
  \footnotesize
  \setlength{\tabcolsep}{4pt}
  \begin{tabular*}{\textwidth}{@{\extracolsep{\fill}}l l rrr rrr@{}}
    \toprule
    & & \multicolumn{3}{c}{$K = 10$} & \multicolumn{3}{c}{$K = 100$} \\
    \cmidrule(lr){3-5} \cmidrule(lr){6-8}
    Segment & Strategy & Diversity & Coverage$^{\dagger}$ & Novelty & Diversity & Coverage$^{\dagger}$ & Novelty \\
    \midrule
    \multirow{4}{*}{All}
    & Centroid          & ---    & ---     & ---     & ---    & ---     & ---     \\
    & TemporalCentroid  & $+0.0$ & $-5.8$  & $+1.9^{\ast}$  & $-0.1^{\ast}$ & $-8.7$  & $+0.5^{\ast}$  \\
    & Narrative         & $+4.2^{\ast}$ & $-79.8$ & $-17.8^{\ast}$ & $+3.0^{\ast}$ & $-82.2$ & $-19.0^{\ast}$ \\
    & TemporalNarrative & $+4.0^{\ast}$ & $-79.2$ & $-19.9^{\ast}$ & $+3.2^{\ast}$ & $-83.7$ & $-20.5^{\ast}$ \\
    \midrule
    \multirow{4}{*}{Non-Explorer}
    & Centroid          & ---    & ---     & ---     & ---    & ---     & ---     \\
    & TemporalCentroid  & $+0.1$ & $-5.9$  & $+1.6^{\ast}$  & $-0.0$ & $-8.7$  & $+0.3^{\ast}$  \\
    & Narrative         & $+4.1^{\ast}$ & $-78.3$ & $-17.5^{\ast}$ & $+3.0^{\ast}$ & $-81.1$ & $-18.7^{\ast}$ \\
    & TemporalNarrative & $+4.0^{\ast}$ & $-77.6$ & $-19.6^{\ast}$ & $+3.2^{\ast}$ & $-82.7$ & $-20.3^{\ast}$ \\
    \midrule
    \multirow{4}{*}{Explorer}
    & Centroid          & ---    & ---     & ---     & ---    & ---     & ---     \\
    & TemporalCentroid  & $-0.1$ & $-4.1$  & $+2.9^{\ast}$  & $-0.2^{\ast}$ & $-5.9$  & $+1.1^{\ast}$  \\
    & Narrative         & $+4.3^{\ast}$ & $-74.0$ & $-18.9^{\ast}$ & $+3.1^{\ast}$ & $-77.2$ & $-20.7^{\ast}$ \\
    & TemporalNarrative & $+3.9^{\ast}$ & $-74.0$ & $-21.1^{\ast}$ & $+3.1^{\ast}$ & $-79.2$ & $-21.6^{\ast}$ \\
    \bottomrule
  \end{tabular*}
\end{table*}

\subsection{Results}
\label{sec:results}

\subsubsection{Segment-conditional accuracy}
\label{sec:results_accuracy}

Results in this section are reported at PCT~=~25\%, a representative value within the stable operating range identified in Section~\ref{sec:results_pct}. Table~\ref{tab:headline} reports accuracy metrics at $K = 10$ and $K = 100$ across the three segments. The headline finding is that the relative ordering of LLM-based and aggregate strategies is conditional on the user segment.

On the All segment, both LLM-based strategies underperform Centroid on Recall@10 by approximately 31\% and 22\% respectively, while TemporalCentroid is essentially indistinguishable from Centroid. The same pattern holds at $K = 100$ and across NDCG and HitRate. Because the All segment is dominated by Non-Explorer users, this aggregate result largely reflects the majority population.

The \textit{Non-Explorer} segment confirms this reading. Narrative trails Centroid by approximately 32\% on Recall@10 and TemporalNarrative by approximately 22\%, while TemporalCentroid is within a few percent of Centroid across metrics. Aggregate profiling, with or without temporal disentanglement, is the stronger choice for habitual users, and LLM-based profiling provides no accuracy benefit.

\begin{figure*}[t]
  \centering
  \includegraphics[width=\linewidth]{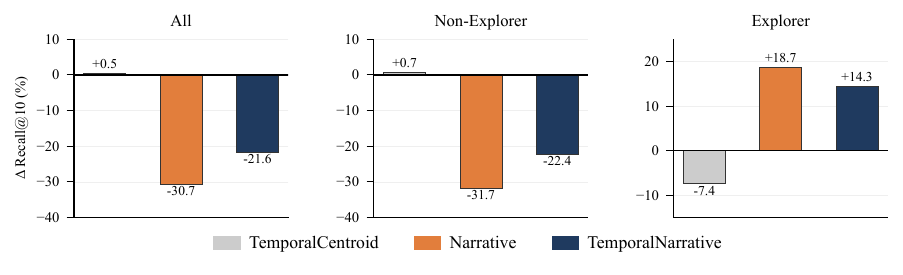}
  \caption{\small Recall@10 across user segments, reported as percentage change relative to the within-segment Centroid baseline at $\text{PCT} = 25\%$. Independent y-axis scale per panel. The relative ordering of LLM-based profiling strategies (Narrative, TemporalNarrative) and aggregate strategies (TemporalCentroid) inverts on the Explorer panel.}
  \label{fig:fig1_segment}
\end{figure*}

The \textit{Explorer} segment produces the inversion. The same LLM-based strategies that underperformed on the majority population now outperform Centroid on Recall@10, with Narrative ahead by approximately 19\% and TemporalNarrative by approximately 14\%, while TemporalCentroid is below Centroid by approximately 7\% on Recall@10, though this difference does not reach statistical significance. The same directional pattern holds across NDCG and HitRate, and at $K = 100$. Figure~\ref{fig:fig1_segment} visualizes this contrast: LLM-based bars extend below zero on All and Non-Explorer, and above zero on Explorer.

The segment-conditional flip is most robustly supported at $K = 100$, where both Narrative and TemporalNarrative significantly outperform Centroid on Recall and NDCG on the \textit{Explorer} segment. HitRate differences on Explorer do not reach significance at either cutoff, and Recall at $K = 10$ similarly does not. The $K = 10$ Explorer results should therefore be read as directionally supporting the more robust $K = 100$ Recall/NDCG pattern, which we attribute to the smaller \textit{Explorer} sample size (1{,}761 users). The corresponding LLM-based comparisons on \textit{All} and \textit{Non-Explorer} reach significance for every metric and cutoff in Table~\ref{tab:headline}.

A secondary observation concerns the temporal-disentanglement axis. In the aggregate regime, adding temporal disentanglement (Centroid $\rightarrow$ TemporalCentroid) produces small and inconsistent effects, with marginal improvements on \textit{Non-Explorers} and marginal degradation on \textit{Explorers}. In the LLM regime, however, Narrative is directionally ahead of TemporalNarrative across all three accuracy metrics on Explorers at $K = 10$, by approximately 4 to 10 percentage points; the pattern weakens at $K = 100$. The temporal split does not produce clear gains over the holistic LLM profile on the segment where LLM-based profiling adds value, particularly at the smaller cutoff most relevant to top-of-list ranking.

\subsubsection{Beyond-accuracy trade-offs}
\label{sec:results_beyond}

\begin{figure*}[t]
  \centering
  \includegraphics[width=\linewidth]{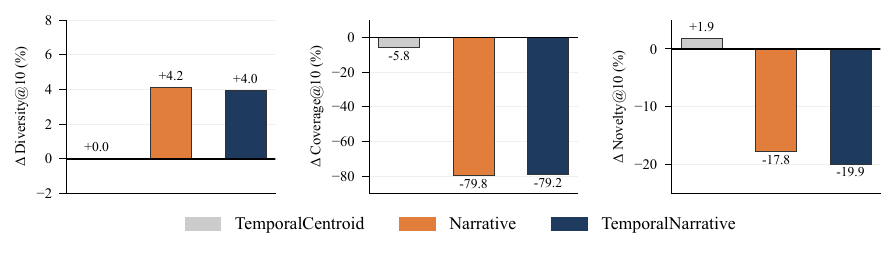}
  \caption{\small Beyond-accuracy metrics at $K = 10$, All segment, $\text{PCT} = 25\%$, reported as percentage change relative to the Centroid baseline. LLM-based profiling produces modestly higher within-list Diversity but substantially lower Coverage and Novelty.}
  \label{fig:fig2_beyond}
\end{figure*}

Beyond accuracy, the four strategies show a distinct pattern across Diversity, Coverage, and Novelty. Figure~\ref{fig:fig2_beyond} reports the comparison at $K = 10$ on the \textit{All} segment.

LLM-based strategies (Narrative, TemporalNarrative) achieve modestly higher within-list Diversity than Centroid, by approximately 4\%. The same strategies post substantially lower Coverage, around -80\%, indicating that recommendations across users are concentrated on a much smaller subset of the catalog. Novelty values follow the same direction, with LLM-based strategies producing recommendations whose self-information is approximately 18-20\% lower than Centroid. TemporalCentroid is within a few percent of Centroid on Diversity and Coverage, with a modest Novelty improvement of approximately 2\%.

The same direction holds across PCT settings and across all three segments. Table~\ref{tab:beyond_accuracy} reports the comparison at $K = 10$ and $K = 100$ for all three segments: LLM-based strategies show modest Diversity gains alongside large Coverage and Novelty losses, while TemporalCentroid is within a few percent of Centroid on all three metrics. The Coverage and Novelty deficits deepen at $K = 100$ relative to $K = 10$, indicating that the effect is not specific to any one cutoff.

Statistical testing supports the popularity-attractor pattern as the most robust empirical signal in our results. Every per-user Novelty comparison between LLM-based and aggregate strategies reaches significance at $\alpha = 0.05$ across all three segments and both cutoffs (Table~\ref{tab:beyond_accuracy}). The corresponding Diversity differences are also significant in every LLM-based-versus-aggregate cell.

\subsubsection{Sensitivity to the short-term window}
\label{sec:results_pct}

\begin{figure*}[t]
  \centering
  \includegraphics[width=\linewidth]{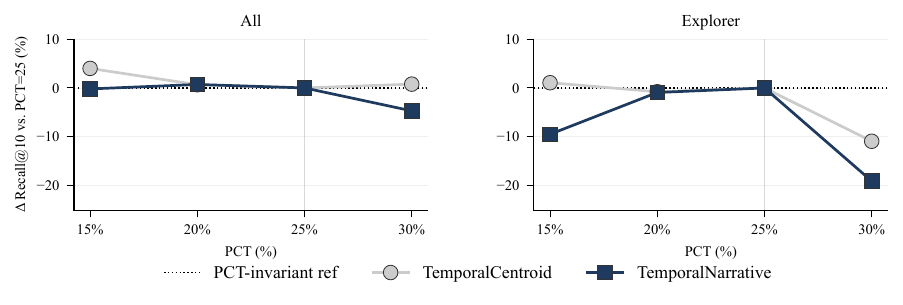}
  \caption{\small Recall@10 sensitivity to the short-term window parameter PCT. Each model's trajectory is normalized to its own value at $\text{PCT} = 25\%$ within the same segment. The dashed line at zero represents the PCT-invariant Centroid and Narrative strategies. TemporalNarrative shows the largest degradation at $\text{PCT} = 30\%$, particularly on the Explorer segment.}
  \label{fig:fig3_pct}
\end{figure*}

Figure~\ref{fig:fig3_pct} reports Recall@10 sensitivity to PCT for the \textit{All} and \textit{Explorer} segments. Centroid and Narrative are PCT-invariant by construction and appear as the dashed zero reference; only TemporalCentroid and TemporalNarrative vary with PCT.

On the \textit{All} segment, both strategies are stable in PCT $\in$ \{15\%, 20\%, 25\%\}, with within-model variation under approximately 4\% either way. At PCT = 30\%, TemporalNarrative drops to approximately $-4.7\%$ while TemporalCentroid remains stable.

The \textit{Explorer} segment shows sharper degradation. TemporalNarrative falls to approximately -9.5\% at PCT = 15\%, recovers between PCT = 20\% and PCT = 25\%, and drops sharply to approximately -19\% at PCT = 30\%. TemporalCentroid is comparatively flat through PCT = \{15\%, 20\%, 25\%\} and drops to approximately -11\% at PCT = 30\%. Both strategies degrade most at PCT = 30\% on \textit{Explorer}, suggesting that the short-term window becomes large enough that the temporal partition stops providing distinguishable signals for users whose recent behavior departs from historical patterns.

Both PCT-sensitive strategies operate stably across PCT values of 15\%, 20\%, and 25\%, and degrade at PCT = 30\%, with the larger drop on the \textit{Explorer} segment. We adopt PCT = 25\% as the default within this stable range for the comparisons in Sections~\ref{sec:results_accuracy} and~\ref{sec:results_beyond}.

\section{Discussion}
\label{sec:discussion}

\subsection{Interpreting the segment-conditional effect}
\label{sec:mechanism}

Why does the relative effectiveness of LLM-based and aggregate profiling flip with user segment? We connect the pattern to the structure of the two representation types.

A Centroid profile is the arithmetic mean of SBERT embeddings of items the user has consumed, and therefore sits near those items in embedding space. When test-period consumption draws from the same neighborhood, this specificity is a strength: ranking against item embeddings efficiently recovers similar items. This is the Non-Explorer regime, where aggregate methods win by a substantial margin.

A Narrative profile is qualitatively different. The LLM compresses the user's history into a natural-language summary that abstracts away from individual items toward themes and preference patterns. When this summary is SBERT-encoded, the resulting vector is positioned in embedding space according to the language of the summary rather than the items that informed it.

For Explorer users, item-level affinity has limited predictive power; what is needed is a representation that captures preference at a level of generality that can extrapolate to new items. Narrative profiles do this by virtue of their thematic abstraction, which is why they outperform Centroid on this segment.

A second mechanistic question concerns the popularity-attractor signature from Section~\ref{sec:results_beyond}. One plausible explanation traces this to training-data bias in the underlying language models. The LLM used for profile generation is trained on web-scale text where popular items are discussed substantially more than long-tail items, producing richer semantic associations for popular content. SBERT, used both for item and profile encoding, is trained on similar text distributions. When user histories are summarized into natural language and encoded into embedding space, the resulting representations gravitate toward regions densely populated by embeddings of popular items, by virtue of the underlying models having more refined semantic coverage of those items. The mechanism predicts the Section~\ref{sec:results_beyond} pattern directly: thematic abstraction raises within-list Diversity while popularity-skewed embedding regions lower Novelty and Coverage. This is consistent with the broader pattern of popularity bias documented in language models~\cite{navigli_2023_biases, gallegos_2024_bias_survey} and in embedding-based retrieval systems~\cite{abdollahpouri_2019_popularity_bias}. While we do not isolate the contributions of the generative and encoding steps individually, the pattern's consistency across PCT settings and segments suggests the effect is structural to LLM-based representation rather than an artifact of specific architectural choices.

A related observation from Section~\ref{sec:results_accuracy} fits the same frame. On \textit{Explorer} at $K = 10$, TemporalNarrative does not improve over the holistic Narrative; at $K = 100$ the two are closer, with neither holding a clear advantage. Temporal disentanglement is informative when there is a coherent recent-versus-historical distinction to disentangle. For \textit{Non-Explorer} users, this distinction exists, and short-term and long-term centroids produce complementary signals. For Explorer users, by construction, recent behavior departs from historical patterns; forcing the LLM to commit to two distinct summaries may decompose a preference structure that does not naturally split along the time axis.

\subsection{Deployment implications}
\label{sec:routing_implications}

The segment-conditional pattern motivates a specific deployment direction: if a sufficiently reliable inference-time signal for exploratory behavior can be constructed, users could be routed to different profiling methods. Habitual users would be served by lower-cost aggregate profiling; exploratory users would be routed to the more expensive LLM-based pipeline where it produces accuracy gains.

The \textit{Explorer}/\textit{Non-Explorer} distinction is post-hoc, computed from the relationship between training and test items. A deployed routing strategy would therefore depend on a separate prediction problem: detecting exploratory behavior from features available at inference time, before the user's near-future consumption is observed. To assess whether such a predictor is plausibly constructible, we examined whether features from training-period behavior correlate with the \textit{Explorer} label using point-biserial correlation~\cite{tate_1954_pointbiserial, hollander_2014_nonparametric}. Two features show significant correlation: history size ($r = -0.30$, $p < 0.001$) and within-history content diversity, measured as the mean pairwise cosine distance of SBERT embeddings across the user's training items ($r = +0.11$, $p < 0.001$). Users with smaller and more diverse training histories are more likely to be \textit{Explorers}. Exploratory behavior is thus partially predictable from training-period features, supporting the feasibility of inference-time routing. The current analysis is a preliminary observation rather than a deployable predictor; substantial work in feature engineering, threshold calibration, and online validation would be needed to construct and deploy such a signal.

A reasonable fallback when an explorer signal is unavailable or noisy is to default to aggregate profiling, which biases the system toward the lower-cost option that is also the stronger choice on the majority population. The routing strategy would also concentrate the popularity-attractor effect (Section~\ref{sec:results_beyond}) on the Explorer population, with attendant lower catalog Coverage and Novelty for those users; production systems would need to weigh this against the accuracy gains.

The routing strategy described here depends on solving the inference-time explorer-detection problem, which we do not address empirically. Our contribution is to demonstrate that segment-conditional heterogeneity exists and is large enough to make selective routing potentially attractive; converting this into a deployable system requires construction and online validation of an explorer-detection signal, and online evaluation of the routing pattern itself.

\subsection{Cost considerations}
\label{sec:cost}

LLM-based profile generation is meaningfully more expensive than aggregate profile construction. Aggregate centroids reduce to mean-pooling over precomputed and cached SBERT embeddings. LLM-based generation requires one or more LLM forward passes per user, with associated compute cost, latency for refresh cycles, and operational complexity in pipeline management.

Selective routing (Section~\ref{sec:routing_implications}) takes this asymmetry into account: by concentrating LLM-based generation on the user population for which it produces accuracy gains, additional compute cost is paid where it produces commensurate value. In our sample, this concentrates LLM-based generation on approximately 18\% of users.

A quantitative cost-benefit analysis would depend on platform-specific factors including subscriber-base size, profile refresh cadence, LLM API pricing or self-hosted compute amortization, and engagement value differences between user segments. We leave that quantification to deployment; the qualitative asymmetry alone makes selective application an attractive design pattern when accuracy gains are themselves selective.

\subsection{Limitations}
\label{sec:limitations}

The evaluation is conducted on a sample drawn from a production streaming platform within a specific content domain. Whether the segment-conditional pattern can be generalized to other platforms, content domains, or LLM-and-encoder pipelines is an open question that warrants replication. Extrapolation to platforms with substantially different user-base characteristics or interaction-pattern distributions should be made cautiously. The evaluation is conducted offline; online deployment may introduce dynamics that this evaluation does not capture.


\section{Conclusion}
\label{sec:conclusion}

We presented a systematic comparison of four semantic user profiling strategies for content-based recommendation, factorially crossed across representation type and temporal handling, evaluated on real-world production data with ranking against the full filtered catalog. The relative effectiveness of aggregate and LLM-based profiling proved to be not uniform between users but conditional on the behavioral regime. Aggregate methods were stronger for habitual users, while LLM-based methods provided accuracy gains on exploratory users whose future consumption diverges from their past behavior. Beyond accuracy, LLM-based profiling exhibited concentration of recommendations on a smaller and more popular slice of the catalog, with measurable reductions in catalog coverage and recommendation novelty. Together, these findings suggest that simpler aggregate methods remain a robust default, while the added cost of LLM-based profiling is justified only selectively, for user segments where it delivers measurable accuracy gains rather than being deployed uniformly across all users.

%% file: main.bbl

\begin{thebibliography}{22}


\ifx \showCODEN    \undefined \def \showCODEN     #1{\unskip}     \fi
\ifx \showDOI      \undefined \def \showDOI       #1{#1}\fi
\ifx \showISBNx    \undefined \def \showISBNx     #1{\unskip}     \fi
\ifx \showISBNxiii \undefined \def \showISBNxiii  #1{\unskip}     \fi
\ifx \showISSN     \undefined \def \showISSN      #1{\unskip}     \fi
\ifx \showLCCN     \undefined \def \showLCCN      #1{\unskip}     \fi
\ifx \shownote     \undefined \def \shownote      #1{#1}          \fi
\ifx \showarticletitle \undefined \def \showarticletitle #1{#1}   \fi
\ifx \showURL      \undefined \def \showURL       {\relax}        \fi
\providecommand\bibfield[2]{#2}
\providecommand\bibinfo[2]{#2}
\providecommand\natexlab[1]{#1}
\providecommand\showeprint[2][]{arXiv:#2}

\bibitem[Abdollahpouri et~al\mbox{.}(2019)]%
        {abdollahpouri_2019_popularity_bias}
\bibfield{author}{\bibinfo{person}{Himan Abdollahpouri}, \bibinfo{person}{Robin Burke}, {and} \bibinfo{person}{Bamshad Mobasher}.} \bibinfo{year}{2019}\natexlab{}.
\newblock \showarticletitle{Managing Popularity Bias in Recommender Systems with Personalized Re-ranking}. In \bibinfo{booktitle}{\emph{Proceedings of the 32nd International Florida Artificial Intelligence Research Society Conference (FLAIRS)}}. \bibinfo{publisher}{AAAI Press}, \bibinfo{pages}{413--418}.
\newblock


\bibitem[Bahdanau et~al\mbox{.}(2014)]%
        {bahdanau2014neural}
\bibfield{author}{\bibinfo{person}{Dzmitry Bahdanau}, \bibinfo{person}{Kyunghyun Cho}, {and} \bibinfo{person}{Yoshua Bengio}.} \bibinfo{year}{2014}\natexlab{}.
\newblock \showarticletitle{Neural Machine Translation by Jointly Learning to Align and Translate}.
\newblock \bibinfo{journal}{\emph{arXiv preprint arXiv:1409.0473}} (\bibinfo{year}{2014}).
\newblock
\showeprint[arxiv]{1409.0473}~[cs.CL]


\bibitem[Castells et~al\mbox{.}(2015)]%
        {castells2015novelty}
\bibfield{author}{\bibinfo{person}{Pablo Castells}, \bibinfo{person}{Neil~J. Hurley}, {and} \bibinfo{person}{Sa{\'u}l Vargas}.} \bibinfo{year}{2015}\natexlab{}.
\newblock \showarticletitle{Novelty and Diversity in Recommender Systems}.
\newblock In \bibinfo{booktitle}{\emph{Recommender Systems Handbook}}, \bibfield{editor}{\bibinfo{person}{Francesco Ricci}, \bibinfo{person}{Lior Rokach}, {and} \bibinfo{person}{Bracha Shapira}} (Eds.). \bibinfo{publisher}{Springer}, \bibinfo{address}{Boston, MA}, Chapter~26.
\newblock
\urldef\tempurl%
\url{https://doi.org/10.1007/978-1-4899-7637-6_26}
\showDOI{\tempurl}


\bibitem[Gallegos et~al\mbox{.}(2024)]%
        {gallegos_2024_bias_survey}
\bibfield{author}{\bibinfo{person}{Isabel~O. Gallegos}, \bibinfo{person}{Ryan~A. Rossi}, \bibinfo{person}{Joe Barrow}, \bibinfo{person}{Md~Mehrab Tanjim}, \bibinfo{person}{Sungchul Kim}, \bibinfo{person}{Franck Dernoncourt}, \bibinfo{person}{Tong Yu}, \bibinfo{person}{Ruiyi Zhang}, {and} \bibinfo{person}{Nesreen~K. Ahmed}.} \bibinfo{year}{2024}\natexlab{}.
\newblock \showarticletitle{Bias and Fairness in Large Language Models: A Survey}.
\newblock \bibinfo{journal}{\emph{Computational Linguistics}} \bibinfo{volume}{50}, \bibinfo{number}{3} (\bibinfo{year}{2024}), \bibinfo{pages}{1097--1179}.
\newblock


\bibitem[Hollander et~al\mbox{.}(2014)]%
        {hollander_2014_nonparametric}
\bibfield{author}{\bibinfo{person}{Myles Hollander}, \bibinfo{person}{Douglas~A. Wolfe}, {and} \bibinfo{person}{Eric Chicken}.} \bibinfo{year}{2014}\natexlab{}.
\newblock \bibinfo{booktitle}{\emph{Nonparametric Statistical Methods} (\bibinfo{edition}{3rd} ed.)}.
\newblock \bibinfo{publisher}{Wiley}, \bibinfo{address}{Hoboken, NJ}.
\newblock
\showISBNx{978-0-470-38737-5}


\bibitem[Kang and McAuley(2018)]%
        {kang2018self}
\bibfield{author}{\bibinfo{person}{Wang-Cheng Kang} {and} \bibinfo{person}{Julian McAuley}.} \bibinfo{year}{2018}\natexlab{}.
\newblock \showarticletitle{Self-attentive sequential recommendation}. In \bibinfo{booktitle}{\emph{2018 IEEE international conference on data mining (ICDM)}}. IEEE, \bibinfo{pages}{197--206}.
\newblock


\bibitem[Kim et~al\mbox{.}(2025)]%
        {kim2025time}
\bibfield{author}{\bibinfo{person}{Yejin Kim}, \bibinfo{person}{Shaghayegh Agah}, \bibinfo{person}{Mayur Nankani}, \bibinfo{person}{Neeraj Sharma}, \bibinfo{person}{Feifei Peng}, \bibinfo{person}{Maria Peifer}, \bibinfo{person}{Sardar Hamidian}, {and} \bibinfo{person}{H~Howie Huang}.} \bibinfo{year}{2025}\natexlab{}.
\newblock \showarticletitle{From Time and Place to Preference: LLM-Driven Geo-Temporal Context in Recommendations}.
\newblock \bibinfo{journal}{\emph{arXiv preprint arXiv:2510.24430}} (\bibinfo{year}{2025}).
\newblock


\bibitem[Kingma and Ba(2014)]%
        {kingma2014adam}
\bibfield{author}{\bibinfo{person}{Diederik~P. Kingma} {and} \bibinfo{person}{Jimmy Ba}.} \bibinfo{year}{2014}\natexlab{}.
\newblock \showarticletitle{Adam: A Method for Stochastic Optimization}.
\newblock \bibinfo{journal}{\emph{arXiv preprint arXiv:1412.6980}} (\bibinfo{year}{2014}).
\newblock
\showeprint[arxiv]{1412.6980}~[cs.LG]


\bibitem[Lops et~al\mbox{.}(2011)]%
        {lops2011content}
\bibfield{author}{\bibinfo{person}{Pasquale Lops}, \bibinfo{person}{Marco De~Gemmis}, {and} \bibinfo{person}{Giovanni Semeraro}.} \bibinfo{year}{2011}\natexlab{}.
\newblock \showarticletitle{Content-based recommender systems: State of the art and trends}.
\newblock \bibinfo{journal}{\emph{Recommender systems handbook}} (\bibinfo{year}{2011}), \bibinfo{pages}{73--105}.
\newblock


\bibitem[Lubos et~al\mbox{.}(2024)]%
        {lubos2024llm}
\bibfield{author}{\bibinfo{person}{Simon Lubos}, \bibinfo{person}{Thi Ngoc~Trang Tran}, \bibinfo{person}{Alexander Felfernig}, \bibinfo{person}{Seda Polat~Erdeniz}, {and} \bibinfo{person}{Viet~Man Le}.} \bibinfo{year}{2024}\natexlab{}.
\newblock \showarticletitle{LLM-Generated Explanations for Recommender Systems}. In \bibinfo{booktitle}{\emph{Adjunct Proceedings of the 32nd ACM Conference on User Modeling, Adaptation and Personalization}}. \bibinfo{publisher}{ACM}, \bibinfo{pages}{276--285}.
\newblock


\bibitem[Ma et~al\mbox{.}(2024)]%
        {ma2024xrec}
\bibfield{author}{\bibinfo{person}{Qiang Ma}, \bibinfo{person}{Xinyi Ren}, {and} \bibinfo{person}{Chao Huang}.} \bibinfo{year}{2024}\natexlab{}.
\newblock \showarticletitle{XRec: Large Language Models for Explainable Recommendation}. In \bibinfo{booktitle}{\emph{Findings of the Association for Computational Linguistics: EMNLP 2024}}. \bibinfo{publisher}{Association for Computational Linguistics}, \bibinfo{pages}{391--402}.
\newblock


\bibitem[Navigli et~al\mbox{.}(2023)]%
        {navigli_2023_biases}
\bibfield{author}{\bibinfo{person}{Roberto Navigli}, \bibinfo{person}{Simone Conia}, {and} \bibinfo{person}{Bj{\"o}rn Ross}.} \bibinfo{year}{2023}\natexlab{}.
\newblock \showarticletitle{Biases in Large Language Models: Origins, Inventory, and Discussion}.
\newblock \bibinfo{journal}{\emph{Journal of Data and Information Quality}} \bibinfo{volume}{15}, \bibinfo{number}{2} (\bibinfo{year}{2023}), \bibinfo{pages}{1--21}.
\newblock
\urldef\tempurl%
\url{https://doi.org/10.1145/3597307}
\showDOI{\tempurl}


\bibitem[Reimers and Gurevych(2019)]%
        {reimers-2019-sentence-bert}
\bibfield{author}{\bibinfo{person}{Nils Reimers} {and} \bibinfo{person}{Iryna Gurevych}.} \bibinfo{year}{2019}\natexlab{}.
\newblock \showarticletitle{Sentence-BERT: Sentence Embeddings using Siamese BERT-Networks}. In \bibinfo{booktitle}{\emph{Proceedings of the 2019 Conference on Empirical Methods in Natural Language Processing}}. \bibinfo{publisher}{Association for Computational Linguistics}.
\newblock
\urldef\tempurl%
\url{https://arxiv.org/abs/1908.10084}
\showURL{%
\tempurl}


\bibitem[Sabouri et~al\mbox{.}(2025)]%
        {sabouri2025towards}
\bibfield{author}{\bibinfo{person}{Milad Sabouri}, \bibinfo{person}{Masoud Mansoury}, \bibinfo{person}{Kun Lin}, {and} \bibinfo{person}{Bamshad Mobasher}.} \bibinfo{year}{2025}\natexlab{}.
\newblock \showarticletitle{Towards Explainable Temporal User Profiling with LLMs}. In \bibinfo{booktitle}{\emph{Adjunct Proceedings of the 33rd ACM Conference on User Modeling, Adaptation and Personalization}}. \bibinfo{pages}{219--227}.
\newblock


\bibitem[Sun et~al\mbox{.}(2019)]%
        {sun2019bert4rec}
\bibfield{author}{\bibinfo{person}{Fei Sun}, \bibinfo{person}{Jun Liu}, \bibinfo{person}{Jian Wu}, \bibinfo{person}{Changhua Pei}, \bibinfo{person}{Xiao Lin}, \bibinfo{person}{Wenwu Ou}, {and} \bibinfo{person}{Peng Jiang}.} \bibinfo{year}{2019}\natexlab{}.
\newblock \showarticletitle{BERT4Rec: Sequential recommendation with bidirectional encoder representations from transformer}. In \bibinfo{booktitle}{\emph{Proceedings of the 28th ACM international conference on information and knowledge management}}. \bibinfo{pages}{1441--1450}.
\newblock


\bibitem[Tan et~al\mbox{.}(2016)]%
        {tan2016improved}
\bibfield{author}{\bibinfo{person}{Yong~Kiam Tan}, \bibinfo{person}{Xinxing Xu}, {and} \bibinfo{person}{Yong Liu}.} \bibinfo{year}{2016}\natexlab{}.
\newblock \showarticletitle{Improved Recurrent Neural Networks for Session-Based Recommendations}. In \bibinfo{booktitle}{\emph{Proceedings of the 1st Workshop on Deep Learning for Recommender Systems}}. \bibinfo{publisher}{ACM}, \bibinfo{pages}{17--22}.
\newblock


\bibitem[Tate(1954)]%
        {tate_1954_pointbiserial}
\bibfield{author}{\bibinfo{person}{Robert~F. Tate}.} \bibinfo{year}{1954}\natexlab{}.
\newblock \showarticletitle{Correlation between a discrete and a continuous variable. Point-biserial correlation}.
\newblock \bibinfo{journal}{\emph{The Annals of Mathematical Statistics}} \bibinfo{volume}{25}, \bibinfo{number}{3} (\bibinfo{year}{1954}), \bibinfo{pages}{603--607}.
\newblock


\bibitem[Vargas and Castells(2011)]%
        {vargas2011rank}
\bibfield{author}{\bibinfo{person}{Sa{\'u}l Vargas} {and} \bibinfo{person}{Pablo Castells}.} \bibinfo{year}{2011}\natexlab{}.
\newblock \showarticletitle{Rank and Relevance in Novelty and Diversity Metrics for Recommender Systems}. In \bibinfo{booktitle}{\emph{Proceedings of the Fifth ACM Conference on Recommender Systems}}. \bibinfo{publisher}{ACM}, \bibinfo{pages}{109--116}.
\newblock


\bibitem[Vaswani et~al\mbox{.}(2017)]%
        {waswani2017attention}
\bibfield{author}{\bibinfo{person}{A Vaswani}, \bibinfo{person}{N Shazeer}, \bibinfo{person}{N Parmar}, \bibinfo{person}{J Uszkoreit}, \bibinfo{person}{L Jones}, \bibinfo{person}{A Gomez}, \bibinfo{person}{L Kaiser}, {and} \bibinfo{person}{I Polosukhin}.} \bibinfo{year}{2017}\natexlab{}.
\newblock \showarticletitle{Attention is all you need}. In \bibinfo{booktitle}{\emph{NIPS}}.
\newblock


\bibitem[Wang et~al\mbox{.}(2019)]%
        {wang2019modeling}
\bibfield{author}{\bibinfo{person}{Chenyang Wang}, \bibinfo{person}{Min Zhang}, \bibinfo{person}{Weizhi Ma}, \bibinfo{person}{Yiqun Liu}, {and} \bibinfo{person}{Shaoping Ma}.} \bibinfo{year}{2019}\natexlab{}.
\newblock \showarticletitle{Modeling Item-Specific Temporal Dynamics of Repeat Consumption for Recommender Systems}. In \bibinfo{booktitle}{\emph{Proceedings of the World Wide Web Conference}}. \bibinfo{publisher}{ACM}, \bibinfo{pages}{1977--1987}.
\newblock


\bibitem[Wilcoxon(1945)]%
        {wilcoxon_1945_individual}
\bibfield{author}{\bibinfo{person}{Frank Wilcoxon}.} \bibinfo{year}{1945}\natexlab{}.
\newblock \showarticletitle{Individual comparisons by ranking methods}.
\newblock \bibinfo{journal}{\emph{Biometrics Bulletin}} \bibinfo{volume}{1}, \bibinfo{number}{6} (\bibinfo{year}{1945}), \bibinfo{pages}{80--83}.
\newblock


\bibitem[Zhu et~al\mbox{.}(2017)]%
        {zhu2017what}
\bibfield{author}{\bibinfo{person}{Yu Zhu}, \bibinfo{person}{Hua Li}, \bibinfo{person}{Yubo Liao}, \bibinfo{person}{Beidou Wang}, \bibinfo{person}{Zhixu Guan}, \bibinfo{person}{Hui Liu}, {and} \bibinfo{person}{Deng Cai}.} \bibinfo{year}{2017}\natexlab{}.
\newblock \showarticletitle{What to Do Next: Modeling User Behaviors by Time-LSTM}. In \bibinfo{booktitle}{\emph{Proceedings of the 26th International Joint Conference on Artificial Intelligence (IJCAI 2017)}}. \bibinfo{publisher}{IJCAI}, \bibinfo{pages}{3602--3608}.
\newblock


\end{thebibliography}
